\documentclass[aps,preprint,nofootinbib,floatfix]{revtex4-1}
\usepackage{graphicx,color}
\usepackage{amsmath,amssymb}
\usepackage{url}
\usepackage{epstopdf}
\newcommand{\lsim}{\mathrel{\mathop{\kern 0pt \rlap
  {\raise.2ex\hbox{$<$}}}
  \lower.9ex\hbox{\kern-.190em $\sim$}}}
\newcommand{\gsim}{\mathrel{\mathop{\kern 0pt \rlap
  {\raise.2ex\hbox{$>$}}}
  \lower.9ex\hbox{\kern-.190em $\sim$}}}

\newcommand{\BM}[1]{{\mbox{\boldmath{$#1$}}}}
\newcommand{\fr}[2]{{\hbox{$ #1 \over #2 $}}}

\begin{document}

\title{Dipole Moment Dark Matter at the LHC}

\author{Vernon Barger$^{1}$, Wai-Yee Keung$^{2}$, Danny Marfatia$^{3}$, Po-Yan Tseng$^{1,4}$}

\affiliation{
$^1$Department of Physics, University of Wisconsin, Madison, WI 53706,
USA \\
$^2$Department of Physics, University of Illinois at Chicago, IL 60607, USA\\
\mbox{$^3$Department of Physics \& Astronomy, University of Kansas, Lawrence, KS 66045, USA\\}
$^4$Department of Physics, National Tsing Hua University, Hsinchu 300,
Taiwan
}


\begin{abstract}

Monojet and monophoton final states with large missing transverse energy (${\not E}_T$) are important for dark matter (DM) searches at colliders. We present analytic expressions for the differential cross sections for the parton-level processes, $q\overline{q}(qg)\rightarrow g(q) \chi \overline{\chi}$ and $q\overline{q}\rightarrow \gamma\chi \overline{\chi}$, for a neutral DM particle with a magnetic dipole moment (MDM) or an electric dipole moment (EDM). We collectively call such DM candidates dipole moment dark matter (DMDM). We also provide monojet cross sections for scalar, vector and axial-vector interactions. 
We then use ATLAS/CMS monojet${+\not E}_T$ data and CMS monophoton$+{\not E}_T$ data to constrain DMDM. 
We find that 7~TeV LHC bounds on the MDM DM-proton scattering cross section are about six orders of magnitude weaker than on the conventional spin-independent cross section.

\end{abstract}

\maketitle

\section{Introduction}


Collider data have provided an important avenue for dark matter (DM) searches, especially for candidates 
lighter than about 10~GeV~\cite{tait,tait2,chi2}, for which direct detection experiments 
have diminished sensitivity due to the small recoil energy of the scattering process.
In fact, current assumption-dependent bounds on spin-dependent DM-nucleon scattering from LHC data, obtained using an effective field theory framework, are comparable or even superior to those from direct detection experiments for DM lighter than a TeV~\cite{tait2,chi2}.   

The final states that have proven to be effective for DM studies at colliders are those with a single jet or single photon
and large missing transverse energy (${\not E}_T$) or transverse momentum. Our goal is study
these signatures for DM that possesses
a magnetic dipole moment (MDM) or an electric dipole moment (EDM)~\cite{kamion}; earlier work can be found in Ref.~\cite{fortin}. Thus, the DM may be a Dirac 
fermion, but not a Majorana fermion. We refer to these DM candidates as dipole moment dark matter (DMDM).
We begin with a derivation of the differential cross sections for the parton-level processes that give 
monojet$+{\not E}_T$ and monophoton$+{\not E}_T$ final states at the LHC.
We then use 7~TeV $j+{\not E}_T$ data from ATLAS~\cite{atlas} and CMS~\cite{cms4.7fb}, and $\gamma+{\not E}_T$ data from CMS~\cite{CMS-monophoton} 
 to constrain DMDM. Finally, we place bounds on the MDM DM-proton scattering cross section.

\section{Production cross sections}

The monojet${+\not E}_T$ and monophoton${+\not E}_T$  final states for DM production at the LHC arise from the $2 \to 3$ parton level processes $q\overline{q}(qg)\rightarrow g(q) \chi \overline{\chi}$ and $q\overline{q}\rightarrow \gamma\chi \overline{\chi}$.
Since the momenta and spin of the final state DM particles can not be measured, their phase space
can be integrated out. Thus, the $2 \to 3$ processes are simplified to $2 \to 2$ processes. We use this fact to
find analytic expressions for the parton-level cross sections by first focusing on the DM pair $\chi\overline\chi$.

A dark matter particle $\chi$ with magnetic dipole moment $\mu_\chi$ interacts with an electromagnetic field
$F_{\mu\nu}$ through the interaction
${\cal L}=\fr12 \mu_\chi \bar\chi \sigma^{\mu\nu} F_{\mu\nu} \chi$.
The corresponding vertex is
${\Gamma_M}^\mu= \bar u(p) i\sigma^{\mu\nu}(p+p')_\nu v(p')$.
Using the Gordon decomposition identity,
$$ \bar u(p) \gamma^\mu v(p') =
\fr{1}{2m_\chi} \bar u(p) [p^\mu- p'^\mu + i\sigma^{\mu\nu} (p+p')_\nu ] v(p') \,,$$
we write ${\Gamma_M}^\mu$ in terms of the 
QED scalar annihilation vertex, ${\Gamma_0}^\mu =(p-p')^\mu$,  and the QED vectorial
vertex for Dirac fermion pair production, ${\Gamma_{1\over2}}^\mu=\bar u(p)\gamma^\mu v(p')$:
$${\Gamma_M}^\mu
= 2m_\chi{\Gamma_{1\over2}}^\mu - {\Gamma_{0}}^\mu \bar u(p) v(p') \ .$$

Consider ${\Gamma_0}^\mu$. Integrating the 2-body phase space,
\begin{equation}
dps_2(P=p+p')=(2\pi)^4 \delta^4(P-p-p')
                            {d^3 \BM p\over (2\pi)^3 2E_{p}}
                            {d^3 \BM p'\over (2\pi)^3 2E_{p'}}  \,, \nonumber
\end{equation}
gives
\begin{equation}
\int dps_2(P=p+p')=\fr{1}{8\pi} \sqrt{1-4m_\chi^2/P^2} \,. \nonumber
\end{equation}
The relevant tensor that enters the calculation of the cross section is
$$ {T_0}^{\mu\nu} \equiv \int {\Gamma_0}^\mu ({\Gamma_0}^\nu)^* dps_2(P=p+p')
 \,.$$
Gauge invariance, $P_\mu {T_0}^{\mu\nu}=0$, dictates that ${T_0}^{\mu\nu}$ take  the  form,
$$ {T_0}^{\mu\nu} =  S_0  (P^2 g^{\mu\nu} -P^\mu P^\nu) \ . $$
{\it i.e.}, $ {{T_0}^\mu}_\mu=3 P^2 S_0 $. Thus to determine $S_0$, we can circumvent the more
involved tensor calculation by simply evaluating
$$ {{T_0}^\mu}_\mu=\int (p-p')^2 dps_2(P=p+p')=
\int (2m_\chi^2-2 p\cdot p') dps_2=-\fr{q^2}{8\pi}
(1-4m_\chi^2/P^2)^{3\over2}    \,$$
$$ \implies {S_0}=-\fr13\fr{1}{8\pi}                   
(1-4m_\chi^2/P^2)^{3\over2}    \ .$$

Now we study ${\Gamma_{1\over2}}^\mu$. By analogy to
${T_0}^{\mu\nu}$, we define ${T_{1\over2}}^{\mu\nu}$ via
$$ {T_{1\over2}}^{\mu\nu} \equiv \sum_{spin} 
\int {\Gamma_{1\over2}}^\mu ({\Gamma_{1\over2}}^\nu)^* dps_2(P=p+p')
=S_{1\over2}  (P^2 g^{\mu\nu} -P^\mu P^\nu) \ . $$
Taking the trace, we get
$$ 3P^2 S_{1\over2}={\rm Tr} \int (\not p+m_\chi)\gamma^\mu (\not p'-m_\chi) \gamma_\mu
dps_2
={\rm Tr} \int (-2 \not p \not p' -4m_\chi^2 \BM 1) dps_2\,$$
$$ \implies S_{1\over2}=-\fr43 \fr{1}{8\pi} (1+2m_\chi^2/P^2)(1-4m_\chi^2/P^2)^{1\over2} \,. $$
In the high energy limit ($P^2 \gg 4m_\chi^2$), $S_{1\over2}=4S_0$, as expected by counting
degrees of freedom.

The corresponding $S_M$ for the MDM case can be obtained from the previous calculations and an additional calculation of the interference term,
$$  -2 (2m_\chi) \hbox{ Tr } (\not p'-m_\chi) \gamma^\mu (\not p +m_\chi) (p-p')_\mu
=-16m_\chi^2 P^2 (1-4m_\chi^2/P^2)\,.$$
We find
$$ {S_M}=4m_\chi^2 S_{1\over2} +2q^2(1-4m_\chi^2/q^2) S_0 + S_X\,,  $$
with $S_X=-\fr{16}{3} \fr{1}{8\pi}m_\chi^2(1-4m_\chi^2/q^2)^{3\over2}$. Therefore,
$$ S_M =
-\fr23 \fr{1}{8\pi} P^2(1+8m_\chi^2/P^2)\sqrt{1-4m_\chi^2/P^2}\,.$$

We are interested in e.g., $q(p_1)+ \bar q(p_2) \to g(p_3)+[\chi\bar \chi](P)$, with $s=(p_1+p_2)^2$, 
$t= (p_1-p_3)^2$, $u=(p_2-p_3)^2$, and $s+t+u=P^2$, the invariant mass squared of the DM pair $\chi\bar\chi$. This defines our notation.
Multiplying the cross sections for Drell-Yan at high $p_T$~\cite{field} by $S_M(m_\chi)/S_{1\over 2}(m_\ell=0)$ (with an appropriate modification of couplings), we obtain 
\begin{equation}
{d\sigma^{MDM}  \over dt dP^2  }(q\bar q \to b [\chi\bar \chi])
={C_b e^2 e_q^2 \over 16\pi s^2} {\mu_\chi^2\over 24\pi^2}{8 \over 9}
 { (t-P^2)^2 +  (u-P^2)^2 \over tu }
\left(1+{8m_\chi^2\over P^2}\right)
\left(1-{4m_\chi^2\over P^2}\right)^{1\over2}\;,
\label{mdm1}
\end{equation}
\begin{equation}
{d\sigma^{MDM} \over dt dP^2  } (q g  \to q [\chi\bar \chi])    
={g_s^2 e^2 e_q^2 \over 16\pi s^2} {\mu_\chi^2\over 24\pi^2}{1\over 3} { (u-P^2)^2 +  (s-P^2)^2 \over -su } 
\left(1+{8m_\chi^2\over P^2}\right)
\left(1-{4m_\chi^2\over P^2}\right)^{1\over2}\;,
\label{mdm2}
\end{equation}
where $e_q$ is the quark charge in units of $e$. If the gauge boson $b$ is a gluon, $C_b=g^2_s$, and if it is a photon, $C_b=\frac{3}{4}e^2_qe^2$.

The interaction Lagrangian for a DM particle with EDM $d_\chi$ is  ${\cal L}=\fr12 d_\chi \bar\chi \sigma^{\mu\nu} \gamma_5 F_{\mu\nu} \chi$.
A similar procedure gives the EDM DM cross sections, 
\begin{equation}
{d\sigma^{EDM}  \over dt dP^2  }(q\bar q \to b [\chi\bar \chi])
={C_b e^2 e_q^2 \over 16\pi s^2} {d_\chi^2\over 24\pi^2} {8 \over 9} { (t-P^2)^2 +  (u-P^2)^2 \over tu } 
\left(1-{4m_\chi^2\over P^2}\right)^{3\over2}\;,
\label{edm1}
\end{equation}
\begin{equation}
{d\sigma^{EDM} \over dt dP^2  } (q g  \to q [\chi\bar \chi])    
={g_s^2 e^2 e_q^2 \over 16\pi s^2}  {d_\chi^2\over 24\pi^2} {1\over 3}{ (u-P^2)^2 +  (s-P^2)^2 \over -su }
\left(1-{4m_\chi^2\over P^2}\right)^{3\over2}\;.
\label{edm2}
\end{equation}

DMDM interacts with the $Z$-boson via  
the relevant  dimension-5 Lagrangian, 
{$ {\cal  L}=\frac{1}{2} \overline{\chi}\sigma^{\mu \nu} (d_B+d_E\gamma_5)
\chi Z_{\mu \nu}$, where 
$Z_{\mu \nu}=\partial_{\mu}Z_{\nu}-\partial_{\nu}Z_{\mu}$.  
The fermion line of the final DM state is
$$ {\Gamma_Z}^{\mu}=\overline{u}(p) \sigma^{{\mu} {\rho}} 
(d_B+d_E \gamma_5) (p+p')_{{\rho}} v(p')\,.$$
On doing the phase space integration, the following tensor appears:
$$ {T_Z}^{\mu \nu}=\sum_{spin} \int {\Gamma_Z}^{\mu} ({\Gamma_Z}^{\nu})^{\dagger} dps_2=
S_Z(P^2g^{\mu \nu}+P^{\mu}P^{\nu})\;.$$
Its trace is
$${{T_Z}^{\mu}}_{\mu}=3P^2S_{Z}=
(-\pi P^4)\frac{1}{(2\pi)^2} \left[ d^2_B\left( 1+\frac{8m_\chi^2}{P^2} \right)+
d^2_E\left( 1-\frac{4m_\chi^2}{P^2} \right)   \right]
\left( 1-\frac{4m_\chi^2}{P^2} \right)^{\frac{1}{2}}\;,$$
$$\implies S_{Z} = -\frac{\pi}{3} P^2\frac{1}{(2\pi)^2} \left[ d^2_B\left( 1+\frac{8m_\chi^2}{P^2} \right)+
d^2_E\left( 1-\frac{4m_\chi^2}{P^2} \right)   \right]
\left( 1-\frac{4m_\chi^2}{P^2} \right)^{\frac{1}{2}}\;.$$
In general, we expect interference from the photon MDM $\mu_\chi$ 
and EDM $d_\chi$ amplitudes.
After integrating out the two-body phase space 
of the final state DM, the differential cross sections are 
\begin{eqnarray}
\frac{d\sigma^{\gamma,Z}}{dtdP^2}
      (q\overline{q} \rightarrow  g[\overline{\chi} \chi]) &=&
\frac{1}{16\pi s^2} 
\frac{ g^2_s e^2}{27\pi^2 } 
 \frac{(P^2-u)^2+(P^2-t)^2}{tu} 
\left( 1-\frac{4m^2_{\chi}}{P^2} \right)^{\frac{1}{2}}  \nonumber  \\
&\times& \sum_{i=E,B}^{\ } 
     \left( 1+\frac{F_i m^2_{\chi}}{P^2} \right) P^4
\left[     \left| \frac{g_A^q d_i}{P^2-M^2_Z+iM_Z\Gamma_Z}   \right|^2
\right. \nonumber\\
&  &       \qquad\qquad  \left. +  \left|  \frac{e_q d_i^\gamma }{P^2}+
                  \frac{g_V^q d_i}{P^2-M^2_Z+iM_Z\Gamma_Z} 
           \right|^2 \right]\,, \label{int1}
\end{eqnarray}
\begin{eqnarray}
\frac{d\sigma^{\gamma,Z}}{dtdP^2}  
   (q g  \rightarrow q[\overline{\chi} \chi]) &=&                   
\frac{1}{16\pi s^2}\frac{ g^2_s e^2}{72\pi^2 }
\frac{(P^2-u)^2+(P^2-s)^2}{-su}
\left( 1-\frac{4m^2_{\chi}}{P^2} \right)^{\frac{1}{2}}  \nonumber  \\ 
&\times& \sum_{i=E,B}^{\ }   
     \left( 1+\frac{F_i m^2_{\chi}}{P^2} \right) P^4           
\left[     \left| \frac{g_A^q d_i}{P^2-M^2_Z+iM_Z\Gamma_Z}   \right|^2 
\right. \nonumber\\
&  &       \qquad\qquad  \left. +  \left|  \frac{e_q d_i^\gamma }{P^2}
+          \frac{g_V^q d_i}{P^2-M^2_Z+iM_Z\Gamma_Z}                    
           \right|^2 \right]\,,            \label{int2}                 \end{eqnarray} 
where we use the notation, $d_B^\gamma\equiv \mu_\chi$ and $d_E^\gamma\equiv d_\chi$, to keep Eqs.~(\ref{int1})
and (\ref{int2}) compact. Here, $F_B=8$, $F_E=-4$, and
$x_W=\sin^2 \vartheta_W \approx 0.23$,
$ g^q_V\sin\vartheta_W\cos\vartheta_W =\frac{1}{2}(T^q_3)_L- e_q \sin^2 \vartheta_W$ and 
$g^q_A \sin\vartheta_W\cos\vartheta_W =-\frac{1}{2}(T^q_3)_L$
define the quark-$Z$ boson couplings. In what follows, we set $d_B=d_E=0$.

For the sake of completeness, we also work out the monojet cross sections for the scalar, vector, and axial-vector interactions. The amplitudes are
$G_{q,0} (\bar q q)(\bar \chi \chi)$, 
$ G_{q,V} (\bar q \gamma_\mu q)(\bar \chi \gamma^\mu \chi)$, and
$ G_{q,A} (\bar q \gamma_\mu \gamma_5  q)(\bar \chi \gamma^\mu \gamma_5 \chi)$, respectively.

For the scalar case,
\begin{equation}
{d\sigma^S  \over dt dP^2  }(q\bar q \to g [\chi\bar \chi])
={g_s^2 G_{q,0}^2\over 16\pi s^2} {P^2 \over 16\pi^2  }
{8\over9 }{ s^2+P^2  \over tu }
\left(1-{4m_\chi^2\over P^2}\right)^{3\over2}\;,
\end{equation}
\begin{equation}
{d\sigma^S \over dt dP^2  } (q g  \to q [\chi\bar \chi])    
={g_s^2 G_{q,0}^2\over 16\pi s^2} {P^2 \over 16\pi^2  }
{1\over3} { t^2+P^2  \over -su }
\left(1-{4m_\chi^2\over P^2}\right)^{3\over2}\;.
\end{equation}
For the vector case,
\begin{equation}
{d\sigma^V  \over dt dP^2  }(q\bar q \to g [\chi\bar \chi])
={g_s^2 G_{q,V}^2\over 16\pi s^2} {P^2 \over 12\pi^2  }
{8\over9} { (t-P^2)^2 +  (u-P^2)^2 \over tu }
\left(1-{4m_\chi^2\over P^2}\right)^{1\over2}
\left(1+{2m_\chi^2\over P^2}\right)\;,
\end{equation}
\begin{equation}
{d\sigma^V  \over dt dP^2  }(qg \to q [\chi\bar \chi])
={g_s^2 G_{q,V}^2\over 16\pi s^2} {P^2 \over 12\pi^2  }
{1\over3} { (s-P^2)^2 +  (u-P^2)^2 \over -su }
\left(1-{4m_\chi^2\over P^2}\right)^{1\over2}
\left(1+{2m_\chi^2\over P^2}\right)\;.
\end{equation}
For the axial-vector case,
\begin{equation}
{d\sigma^{AV}  \over dt dP^2  }(q\bar q \to g [\chi\bar \chi])
={g_s^2 G_{q,A}^2\over 16\pi s^2} {P^2 \over 12\pi^2  }
{8\over9} { (t-P^2)^2 +  (u-P^2)^2 \over tu }
\left(1-{4m_\chi^2\over P^2}\right)^{3\over2}\;,
\end{equation}
\begin{equation}
{d\sigma ^{AV} \over dt dP^2  }(qg \to q [\chi\bar \chi])
={g_s^2 G_{q,A}^2\over 16\pi s^2} {P^2 \over 12\pi^2  }
{1\over3} { (s-P^2)^2 +  (u-P^2)^2 \over -su }
\left(1-{4m_\chi^2\over P^2}\right)^{3\over2}\;.
\end{equation}
The kinematic limits for the subprocess are
$P^2 \in [(2m_\chi)^2, s]$, $-t \in [0, s-P^2] $.
For $\not p_T$ cuts, there are additional kinematic constraints.

The above equations apply for Dirac fermion DM. For Majorana DM, there are only scalar and axial-vector interactions. All the other interactions are absent. The results for Majorana DM can be obtained from the corresponding equations by dividing by $2$ (since the 2-body phase space for two identical particles is half that for two distinct particles).

\section{Constraints}

The vertices defining DMDM interactions with the electromagnetic field are
\begin{equation}
V_{\gamma \chi \bar{\chi}}(MDM)= \frac{e}{{\Lambda}_{MDM}} {\sigma}^{\mu \alpha} P_{\mu}\;, \nonumber
\end{equation}
\begin{equation}
V_{\gamma \chi \bar{\chi}}(EDM)= \frac{e}{{\Lambda}_{EDM}} {\sigma}^{\mu \alpha} P_{\mu} \gamma_5\;, \nonumber
\end{equation}
where $P$ is the photon's 4-momentum vector and $\alpha$ is the Dirac index of the photon field. The effective cutoff scales $\Lambda_{MDM}$ and $\Lambda_{EDM}$ are defined so that $\mu_\chi=e/\Lambda_{MDM}$ and 
$d_\chi=e/\Lambda_{EDM}$, in order to facilitate comparison.
They may be related to compositeness or short distance physics, but are not necessarily new physics scales. 

Since monojet$+{\not E}_T$ data from ATLAS and 
CMS~\cite{atlas,cms4.7fb}, and monophoton$+{\not E}_T$ data from CMS~\cite{CMS-monophoton}, at the 7~TeV LHC, are consistent with the SM, we may
use these data to constrain the DMDM cutoff scales.
From an analysis of 1/fb of monojet data, with the requirement that the hardest jet have $p_T>350$~GeV, or $p_T>250$~GeV, or $p_T>120$~GeV, and pseudorapidity $|\eta|<2$,
the ATLAS collaboration has placed $95\%$ C.L. upper limits on the production cross section of 0.035~pb, 0.11~pb
and 1.7~pb, respectively~\cite{atlas}. 
In 5/fb of data, CMS has observed 1142 monojet events with leading jet $p_T>350$~GeV and $|\eta|<2.4$~\cite{cms4.7fb}, to be compared with the standard model (SM) expectation, $N_{SM}\pm\sigma_{SM}=1225\pm101$. We will calculate both observed and expected $95\%$ C.L. upper limits from CMS monojet data. Using 5/fb data, CMS has searched for the $\gamma+{\not E}_T$ final state with photon $p_T>145$~GeV and $|\eta|<1.44$, and set a $90\%$ C.L. upper limit on the production cross section of about 0.0143~pb~\cite{CMS-monophoton}.

%


\begin{figure}[t!]
\centering
\includegraphics[angle=270,width=3.2in]{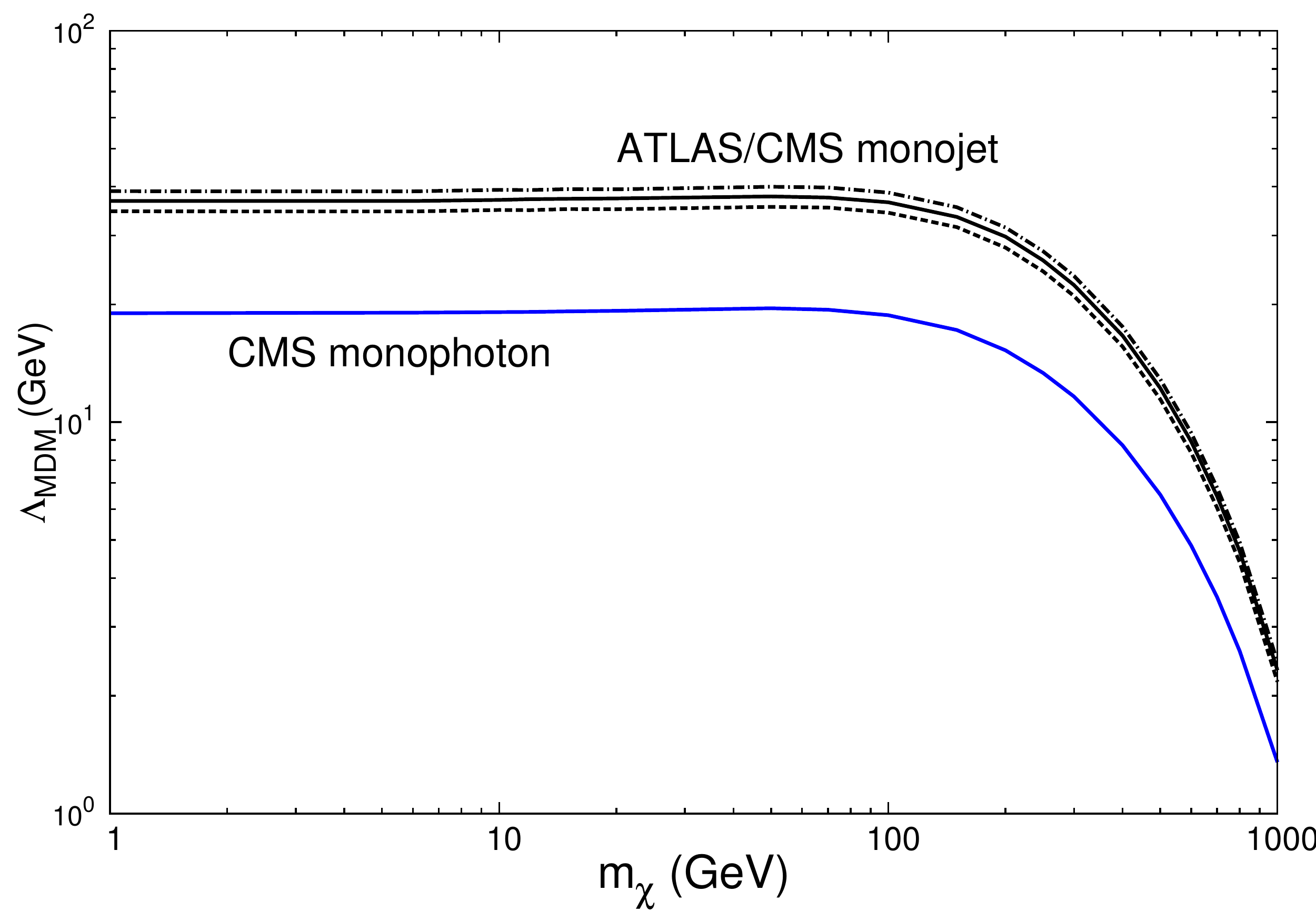}
\includegraphics[angle=270,width=3.2in]{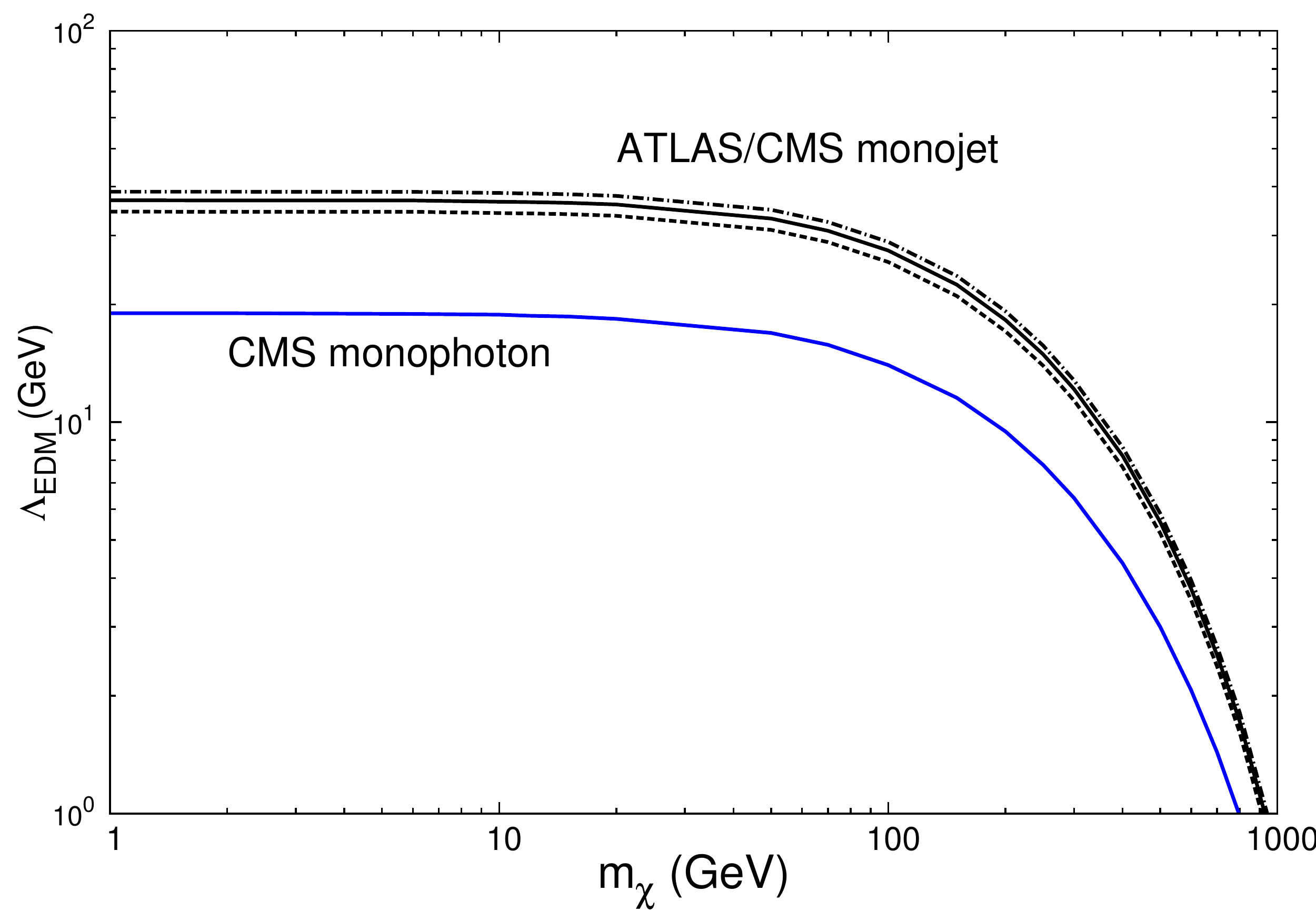}
\caption{\small \label{normal}
The black lines are the $95\%$ C.L. lower limits on the cutoff sales from ATLAS  (solid) and CMS (dash-dotted: observed, dashed: expected) monojet data with leading jet $p_T>350$~GeV and $|\eta|<2$ for ATLAS and $|\eta|<2.4$ for CMS, and the solid blue lines are the $90\%$ C.L. lower limits from the CMS monophoton data. }\end{figure}

To place constraints using the total event rate,
we calculate the cross sections relevant to each detector, ${\sigma}_{ATLAS}$ and ${\sigma}_{CMS}$, of the processes $q\bar{q} \rightarrow g\chi\bar{\chi}$, $qg \rightarrow q\chi\bar{\chi}$ and $q\bar{q} \rightarrow \gamma \chi\bar{\chi}$, by convolving Eqs.~(\ref{mdm1})-(\ref{edm2}) with the
parton distribution functions from CTEQ6~\cite{cteq}. 
For MDM DM,  we have checked that we get the same results from a calculation that begins with an evaluation
of the amplitude squared and the 3-body phase space.
Using CMS $j+{\not E}_T$ data, we place $95\%$ C.L. lower limits on the cutoff scales by requiring~\cite{chi2}
$$ \chi^2 \equiv \frac{[\bigtriangleup_N-N_{DM}(m_{\chi},\Lambda)]^2}{N_{DM}(m_{\chi},\Lambda)+N_{SM}+\sigma^2_{SM} }=3.84\;, $$
where~\cite{cms4.7fb}
$$ \bigtriangleup_N=\left\lbrace  \begin{array}{ll}
200 & \ \ \rm{expected\; bound} \\
158 & \ \ \rm{observed\; bound}\;, \\
\end{array} \right. $$
and $N_{DM}(m_{\chi},\Lambda)={\sigma}_{CMS}\times$luminosity.
The above-mentioned bounds on the production cross sections obtained by the ATLAS and CMS collaborations from the $j+{\not E}_T$ and $\gamma+{\not E}_T$ final states can be used directly to constrain the cutoff scales. 
Figure~\ref{normal} shows lower limits on ${\Lambda}_{MDM}$ and ${\Lambda}_{EDM}$; the
bound from ATLAS corresponds to the $p_T>350$~GeV cut on the hardest jet. We see that for $m_\chi<100$~GeV,
the 95\% C.L. lower limit on the cutoff scales is only about 35~GeV. 
For conventional spin-independent (SI) amplitudes of dimension-6, e.g., 
\begin{equation}
(\overline{q}\gamma_{\mu} q)(\overline{\chi}\gamma^{\mu} \chi)/\Lambda_{SI}^2\,, \ \ \ q=u,d\
\label{vec}
\end{equation}
 typical bounds on $\Lambda_{SI}$ are a few hundred GeV for $m_\chi<100$~GeV, as shown in Fig.~\ref{normal1}.
 The result is counterintuitive since we naively expect the lower limit on $\Lambda_{MDM}$ and $\Lambda_{EDM}$ 
 to be stronger than on $\Lambda_{SI}$ since the DMDM operators are dimension-5. 
 We now explain this result.
 
 Consider MDM DM and the amplitude of Eq.~(\ref{vec}). Neglecting $m_\chi$, and evaluating the cross sections at the peak of the product of the phase space and 
 PDFs for a chosen ${p}_T$ cut, we find
 \begin{equation}
 \frac{\sigma^{SI}(pp\rightarrow j+{\not E}_T)}{\sigma^{MDM}(pp\rightarrow j+{\not E}_T)}
\approx
\frac{8{p}_T^2\Lambda^2_{MDM}}{e^4\Lambda^4_{SI}}\,. \nonumber
 \end{equation}
The left hand side of the equation is unity for an experimental upper bound on the cross section. Then, the lower bound
on $\Lambda_{MDM}$ for a known lower bound on $\Lambda_{SI}$ is $e^2 \Lambda_{SI}^2/(2\sqrt{2} p_T)$. From
Fig.~\ref{normal1}, the 95\% C.L. lower limit on $\Lambda_{SI}$ is 700~GeV for a ${p}_T$ cut of 350~GeV, which translates into a 95\% C.L. lower limit on $\Lambda_{MDM}$ of 45~GeV.


\begin{figure}[t!]
\centering
\includegraphics[angle=270,width=4.5in]{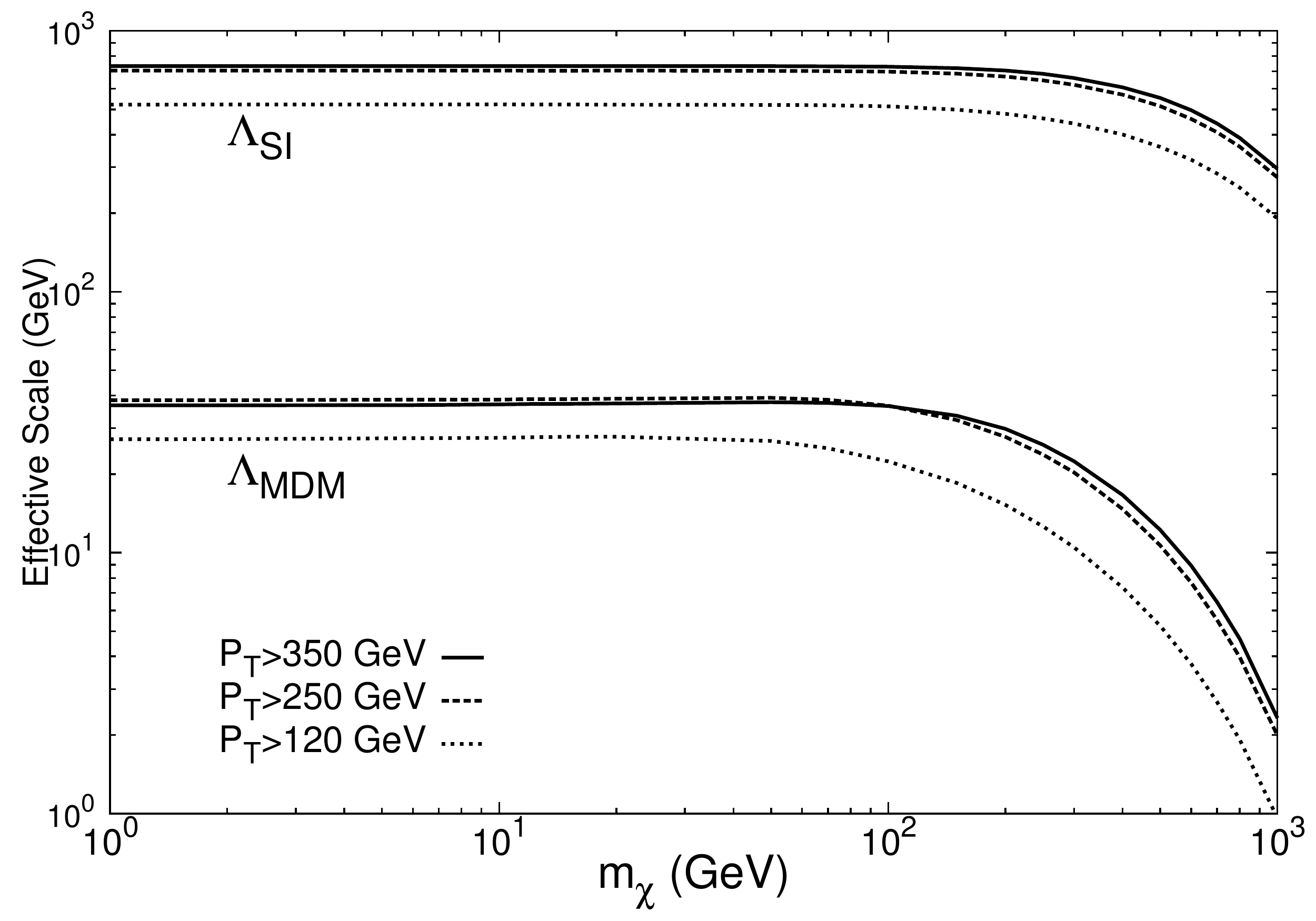}
\caption{\small \label{normal1}
$95\%$ C.L. lower limits from ATLAS $j+{\not E}_T$ data on $\Lambda_{SI}$ and $\Lambda_{MDM}$.}
\end{figure}

\section{Scattering cross sections}

\begin{figure}[t!]
\centering
\includegraphics[angle=270,width=4.5in]{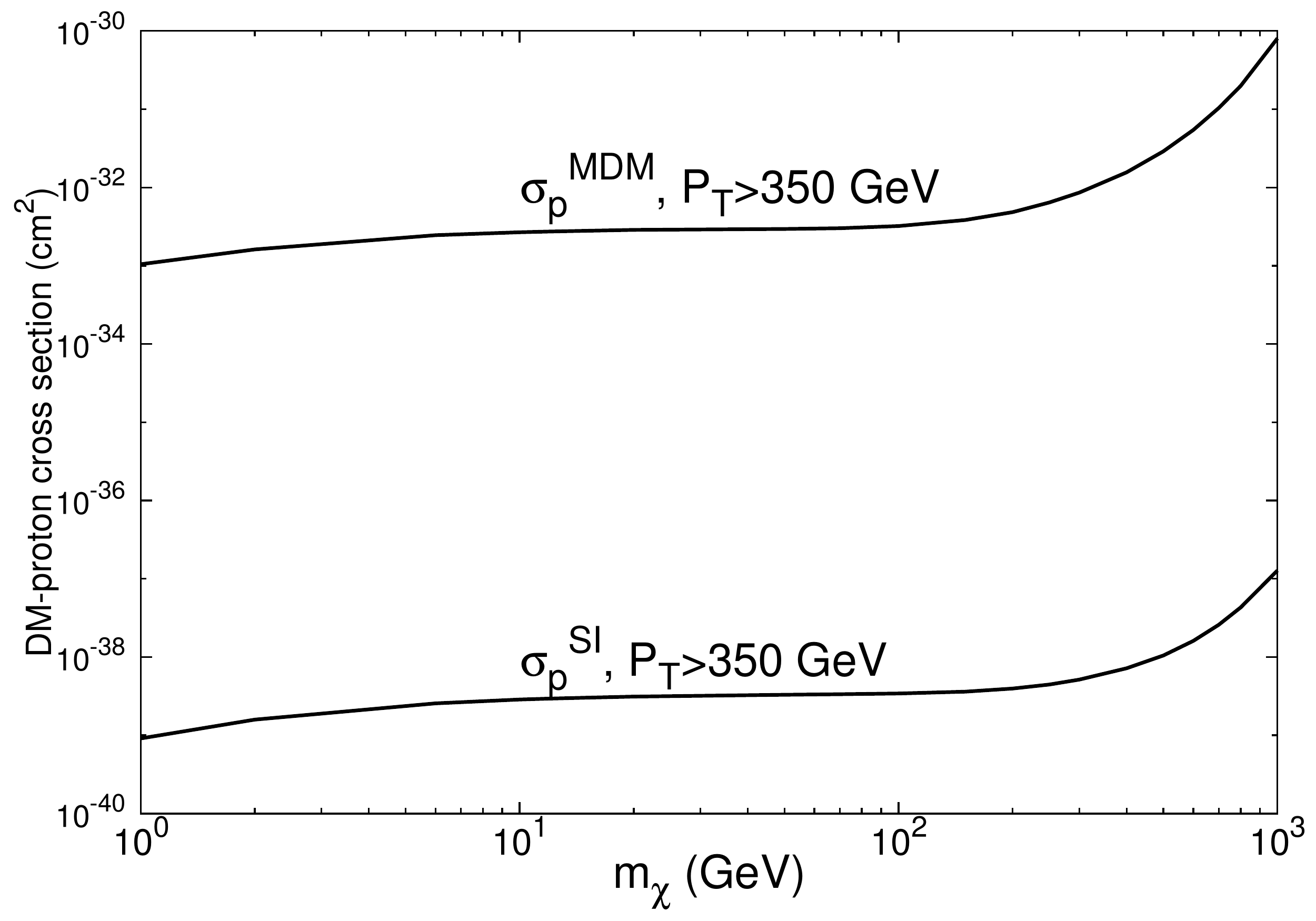}
\caption{\small \label{normal2}
$95\%$ C.L. upper limits on the conventional SI and MDM DM-proton cross sections from ATLAS $j+{\not E}_T$ data.}
\end{figure}

Including the SI and spin-dependent contributions, and setting the electric and magnetic form factors to unity, the MDM DM-proton cross section is~\cite{raby,MDM}{\footnote{The total cross section is
divergent since the Coulomb interaction is singular. Here, we use the energy transfer cross section~\cite{raby} that 
is the same as the usual total cross section for constant differential cross sections.}}
\begin{equation}
{\sigma}^{MDM}_p=\frac{e^4}{2 \pi \Lambda^2_{MDM}}  \left(1-\frac{m^2_{r}}{2m^2_p}-\frac{m^2_{r}}{m_pm_{\chi}}
+\left(\frac{{\mu}_p}{\frac{e}{2m_p}} \right)^2 \frac{m^2_{r}}{m^2_p} 
\right)\;, \nonumber
\end{equation}
where $m_{r}=\frac{m_{\chi}m_p}{m_{\chi}+m_p}$ is the reduced mass of the DM-proton system,
and $\mu_p=2.793e/(2m_p)$ is the MDM of the proton~\cite{webelements}. 
We employ the $95\%$~C.L. lower limit  on $\Lambda_{MDM}$ obtained in Fig.~\ref{normal} from ATLAS data, 
to determine the $95\%$ C.L. upper limit on the MDM DM-proton cross section $\sigma^{MDM}_p$.
This is shown in Fig.~\ref{normal2}.

We now relate limits from the $j+{\not E}_T$ final state on the 
MDM DM-proton scattering cross section to limits on the conventional SI DM-proton cross section.
The DM-proton scattering cross section for the amplitude of Eq.~(\ref{vec}) is
\begin{equation}
\sigma^{SI}_p=\frac{9 m^2_{r}}{\pi\Lambda_{SI}^4}\,. \nonumber
\end{equation}

The $95\%$ C.L. upper limit on $\sigma^{SI}_p$ from ATLAS data is shown in Fig.~\ref{normal2}.
Note that the constraint on $\sigma^{SI}_p$ is about six orders of magnitude more stringent than on 
${\sigma}^{MDM}_p$. This is evident from
\begin{equation}
\frac{\sigma^{SI}_p}{\sigma^{MDM}_p}
\approx \frac{2m^2_p\Lambda^2_{MDM}}{e^4\Lambda^4_{SI}}\,,\nonumber
\label{rat}
\end{equation}
with the limits on $\Lambda_{MDM}$ and $\Lambda_{SI}$ from Fig.~\ref{normal1}. 

The CoGeNT event excess~\cite{CoGeNT} can be explained by a 7~GeV DM particle with a MDM with 
$\Lambda_{MDM}=3$~TeV~\cite{MDM}.
In fact, this candidate can also explain the signals seen by the DAMA~\cite{dama} and 
CRESST~\cite{CRESST} experiments,
and may survive conservative bounds from other direct detection experiments~\cite{DelNobile:2012tx}. 
From Fig.~\ref{normal}, we conclude that LHC bounds are far from ruling out this candidate. This is in contrast to conventional SI scattering, which for light DM, finds strong constraints in collider experiments.

\section*{Acknowledgments}
This work was supported by the DoE under Grant Nos. DE-FG02-12ER41811, 
DE-FG02-95ER40896 and DE-FG02-04ER41308, by the NSF under
Grant No. PHY-0544278, by the National Science Council of Taiwan under Grant No. 100-2917-I-007-002, 
and by the WARF.


\end{document}